\documentclass[%
 aps,
prl,
superscriptaddress,
 amsmath,amssymb,
reprint,%
]{revtex4-2}

\usepackage{graphicx}
\usepackage{dcolumn}
\usepackage{bm}
\usepackage[utf8]{inputenc}
\usepackage[T1]{fontenc}
\usepackage{mathptmx}
\usepackage{xcolor}

\begin{document}

\preprint{V3}

\title[Antiferromagnetic Hysteresis above the Spin Flop Field]{Antiferromagnetic Hysteresis above the Spin Flop Field}

\author{M. J. Grzybowski}
 \email{Michal.Grzybowski@fuw.edu.pl}
  \affiliation{Department of Applied Physics, Eindhoven University of Technology, PO Box 513, 5600 MB Eindhoven, The Netherlands}
  \affiliation{Faculty of Physics, University of Warsaw, Pasteura 5, Warsaw, Poland}
\author{C. F. Schippers}
 \affiliation{Department of Applied Physics, Eindhoven University of Technology, PO Box 513, 5600 MB Eindhoven, The Netherlands}
%
\author{O.~Gomonay}
 \affiliation{Institute of Physics, Johannes Gutenberg-University Mainz, 55128 Mainz, Germany}
\author{K.~Rubi}
\author{M.~E.~Bal}
\author{U.~Zeitler}
 \affiliation{High Field Magnet Laboratory (HFML-EMFL), Radboud University, 6525 ED Nijmegen, The Netherlands}
\author{A.~Kozioł-Rachwał}
\author{M.~Szpytma}
\author{W.~Janus}
 \affiliation{Faculty of Physics and Applied Computer Science, AGH University of Science and Technology, 30-059 Kraków, Poland}
\author{B.~Kurowska}
\author{S.~Kret}
 \affiliation{Institute of Physics, Polish Academy of Sciences, Lotników 32/46, Warsaw, Poland}
\author{M.~Gryglas-Borysiewicz}
 \affiliation{Faculty of Physics, University of Warsaw, Pasteura 5, Warsaw, Poland}
\author{B.~Koopmans}
\author{H.~J.~M.~Swagten}
 \affiliation{Department of Applied Physics, Eindhoven University of Technology, PO Box 513, 5600 MB Eindhoven, The Netherlands}



\date{\today}

\begin{abstract}
Magnetocrystalline anisotropy is essential in the physics of antiferromagnets and commonly treated as a constant, not depending on an external magnetic field. However, we demonstrate that in CoO the anisotropy should necessarily depend on the magnetic field, which is shown by the spin Hall magnetoresistance of the CoO~|~Pt device. Below the Néel temperature CoO reveals a spin-flop transition at 240 K at 7.0 T, above which a hysteresis in the angular dependence of magnetoresistance unexpectedly persists up to 30 T. This behavior is shown to agree with the presence of the unquenched orbital momentum, which can play an important role in antiferromagnetic spintronics.
\end{abstract}

\maketitle

Antiferromagnetic thin-film materials have attracted a lot of attention recently due to their unique properties that create potential applications in spintronics such as data storage \cite{Wadley2016Science, Jungwirth2016NN} or long-distance spin transport \cite{Lebrun2018Nature, Han2020NN}. Their robustness against external magnetic fields is an important promise for using these antiferromagnetic materials. Therefore, they demand strong magnetic fields for reorientation of the spins, and the understanding of the impact of such fields is essential. The behavior of an antiferromagnet (AF) in high magnetic fields is governed by the competition between magneto-crystalline anisotropy and antiferromagnetic exchange interactions. Detailed knowledge of this anisotropy is key to determine possible magnetic configurations of an antiferromagnet as well as proper device design and experimental geometry for potential applications. It is common and well-documented to treat the anisotropy as a physical parameter that is strictly constant for a given temperature \cite{Hoogeboom2017APL, Baldrati2018PRB, Geprags2020JAP}.

In this letter, we demonstrate that the magnetic anisotropy of a thin-film CoO can be modified by strong magnetic fields. It manifests itself in the angular dependent magnetoresistance of the CoO$|$Pt as clear, reproducible and abrupt changes of resitivity coexisting with the hysteresis around the hard axis, which are present up to the highest tested magnetic fields of $B=30\,\text{T}$. Such behavior cannot be reproduced by a simple macrospin model nor a domain model. We explain it by the contribution of the unquenched orbital momentum into the anisotropy.

We start with an experimental study of spin reorientation induced by the strong magnetic fields in CoO with an adjacent Pt layer. When the magnetic field $\mathbf{B}$ is applied to an AF along an easy axis, the N\'eel vector $\mathbf{n}$, which is the difference of the two sublattice magnetizations, may reorient perpendicularly to the field. It is a well-known spin-flop transition and occurs at the spin-flop field $B_\text{sf}$, which depends on the magnetic anisotropy. This reorientation of $\mathbf{n}$ can be detected by the spin Hall magnetoresistance (SMR) \cite{Fischer2018PRB, Geprags2020JAP, Baldrati2018PRB, Hoogeboom2017APL, Baldrati2020PRL}. The angular dependence of the SMR allows extracting the spin-flop field values for different geometrical configurations of the magnetic field and thus, the intriguing behavior of the anisotropy in CoO can be observed. Here, we show that the spin-flop transition is clearly detected together with a remarkable presence of the hysteretic behavior of the AF that cannot be explained by a constant anisotropy field commonly applied for antiferromagnets. This can be understood as a magnetic anisotropy that should necessarily depend on the external magnetic field magnitude and orientation.

\begin{figure}
\includegraphics[width=0.85\linewidth]{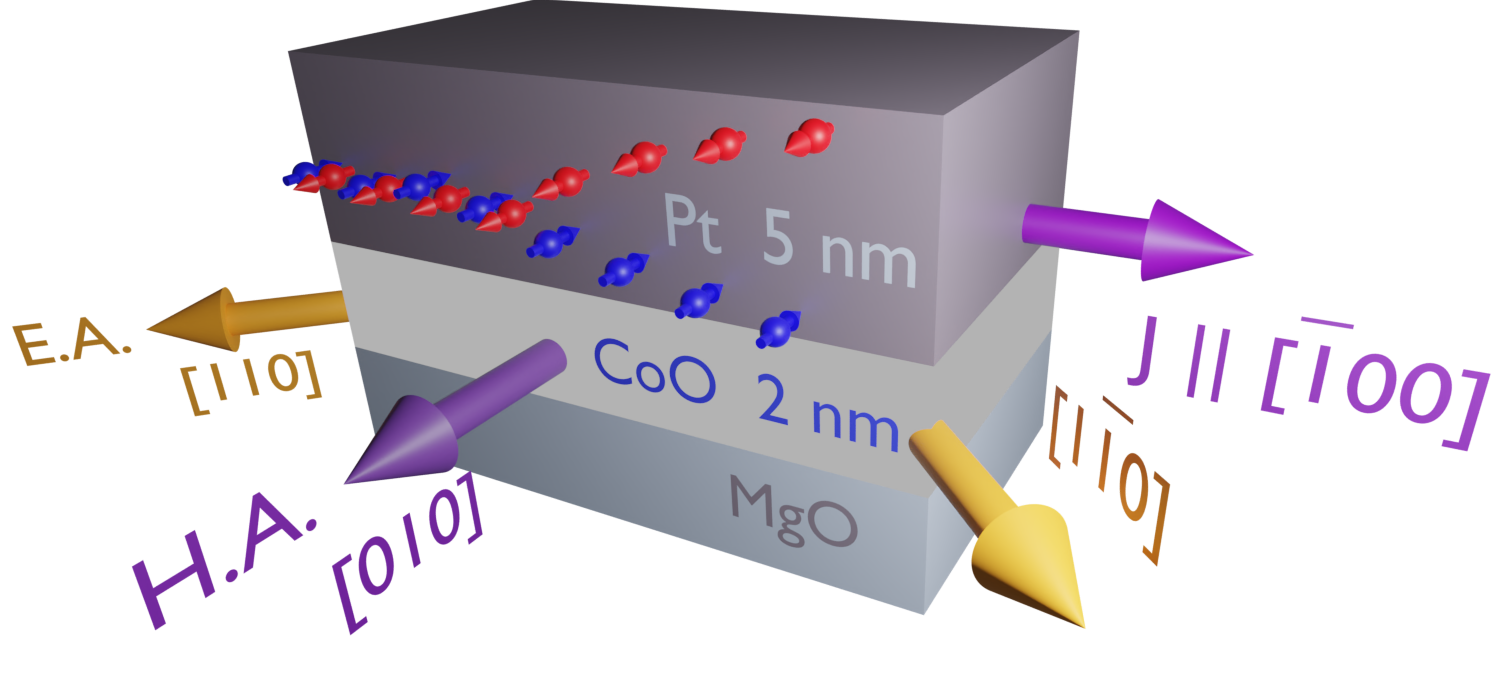}
\caption{\label{fig:hallbar}The scheme of the structure used for the experiments. $\bf{J}$ represents the current direction. Magnetic easy (hard) axes are marked by orange (violet) arrows. Blue and red arrows represent the spin accumulation in Pt due to the spin Hall effect.}
\end{figure}
The electrical measurements were performed on a Hall bar devices made of a CoO thin film grown on MgO (001) single crystals and capped with~5 nm Pt layer. In the main text we present data for the 2~nm CoO prepared by reactive DC magnetron sputtering (Fig.~\ref{fig:hallbar}). For validation purposes, other MBE grown CoO layers, which yielded quantitatively similar results \cite{supplement}, were investigated. Similarly, a reference poly-crystalline 5 nm Pt layer was sputtered directly on MgO and examined which is presented in the supplementary part \cite{supplement}.

We focus on the analysis of the transverse resistance $R_{\text{xy}}$ variations as it can be measured with better accuracy due to the smaller zero-field offset and smaller temperature sensitivity than the longitudinal resistance. The longitudinal data was also analyzed and can be consistently described within the same model \cite{supplement}. Moreover, a reference device made of MgO$|$Pt grown and fabricated in an analogous manner has been studied to rule out effects originating from the mere Pt layer \cite{supplement} such as the Hanle magnetoresistance \cite{Velez2016}. All data presented here were collected below the N\'eel temperature at $T=240\,\text{K}$. The background signal coming likely from small sample misalignment with respect to the magnetic field was subtracted \cite{supplement}.

\begin{figure}
\includegraphics[width=1\linewidth]{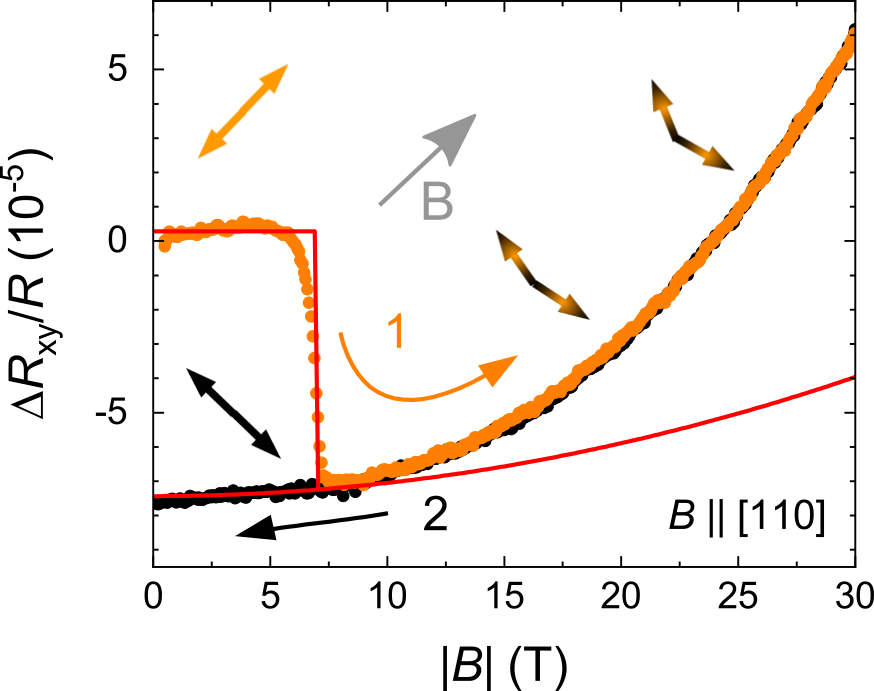}
\caption{\label{fig:spinflopsminus}Transverse magnetoresistance for the CoO structure collected for $B$ along the $[110]$ easy axis. Orange (black) data points correspond to the experiment with increasing (decreasing) magnetic field. The order of the measurements is indicated by the digits. The expected shape of the SMR signal calculated within the macrospin model for $B_{\text{sf}}=7.0\,\text{T}$ is presented by the red curve. The thick arrows represent the directions of the spins (spin canting, angle not to scale) with respect to the magnetic field (thin grey arrow). The electrical signal is expressed as $\Delta R_{\text{xy}}/R$, where $\Delta R_{\text{xy}}$ is the difference in resistance between a data point taken in a magnetic field and the initial value $R_{\text{xy}}(B=0)$. Sheet resistance averaged over different magnetic fields is denoted by $R$.}
\end{figure}

Initially, the expected in-plane biaxial anisotropy of the CoO thin film with easy axes along $[110]$ and $[1\overline{1}0]$ (orange arrows in Fig.~\ref{fig:hallbar}) is verified \cite{Baldrati2020PRL, CaO2011APL}. To do this, the in-plane magnetic field of $30\,\text{T}$ is applied along the expected easy axis $[1\overline{1}0]$ to set a well-defined AF state of $\mathbf{n} \parallel [110]$. Afterwards, the actual field sweep from 0 to $30\,\text{T}$ is performed with the fixed orientation of $\mathbf{B} \parallel [110]$ and the resistance is measured (Fig.~\ref{fig:spinflopsminus}). Upon increasing the magnitude of $\mathbf{B}$ the signal is initially constant and then an abrupt change of $R_\text{xy}$ can be seen at $B=7\,$T. This significant change appears only when the magnitude of $B$ is increasing. Similar behavior is observed with the magnetic field pointing along another easy axis (see \cite{supplement}). The abrupt change of $R_\text{xy}$ does not appear above the N\'eel temperature \cite{supplement}. Moreover, it does not correlate with any random fluctuations of temperature. Finally, the probing current density does not change during the experiments, so thermal effects can be ruled out. Therefore, the electrical signal can be attributed to the transverse SMR. The abrupt change of resistance at $7\,\text{T}$ reflects the spin-flop transition, where the N\'eel vector $\bf{n}$ changes the orientation from $\bf{B} \parallel \bf{n}$ to $\bf{B} \perp \bf{n}$ when $B$ is increasing. The resistance change at the spin flop is consistent with the negative sign of SMR \cite{Baldrati2020PRL, Nakayama2013PRL, Hoogeboom2017APL}. Nonlinear behavior of SMR in high fields can originate from progressive spin canting and the occurrence of small net magnetic moment $\bf{m}$ as depicted symbolically by double-color arrows in Fig.~\ref{fig:spinflopsminus}. 

To verify the qualitative understanding of the obtained results, we construct a macrospin model. Two magnetic moments coupled antiferromagnetically are considered. The energy expression \cite{Gomonay2010PRB} can be formulated as follows:
\begin{equation}
E \left(\bf{n}\right)=\frac{2}{B_{\text{sf}}^2}\left(\bf{B}\cdot\bf{n}\right)^2-\left(n_x^4+n_y^4\right)+\frac{B_{\text{an}\parallel}}{B_{\text{an}\perp}}n_z^2 . \label{eqn:1}
\end{equation}
 In-plane biaxial anisotropy field and out-of-plane uniaxial anisotropy field are represented by $B_{\text{an}\perp}$ and $B_{\text{an}\parallel}$, respectively. The two orthogonal easy directions in-plane are denoted as $x,y$ and correspond to $[110]$ and $[1\overline{1}0]$, respectively (Fig.~\ref{fig:hallbar}). The spin-flop field is denoted by $B_\text{sf}$ and equals $B_\text{sf}=2\sqrt{B_\text{ex}B_{\text{an}\perp}}$ \cite{Gomonay2010PRB}, where $B_{\text{ex}}$ is the exchange field. The value of $B_\text{sf}$ is determined from the experimental data (Fig.~\ref{fig:spinflopsminus}).  $\bf{B}$ remains in the plane of the sample in all experiments. The spins are expected to remain in-plane due to thin-film character of the sample \cite{Baldrati2020PRL, CaO2011APL} and the out-of-plane anisotropy field is set $B_{\text{an}\parallel} \gg B_{\text{an}\perp}$. We perform the energy minimization with $B_{\text{sf}}$ and coordinates of $\mathbf{B}$ as the only parameters, from which we obtain the expected orientation of $\mathbf{n}$. With the assumption of $B \ll B_\text{ex}$, the magnitude of finite $\bf{m}$ perpendicular to $\bf{n}$ due to the tilt of the spin sublattices can be estimated \cite{Gomonay2010PRB, supplement}. The SMR signal can be modeled by $R_{\text{xy}} \propto \rho_n n_\parallel n_\perp+\rho_m m_\parallel m_\perp$. For qualitative analysis, it is enough to approximate the ratio between $\rho_n$ and $\rho_m$ \cite{supplement}. Even when allowing for the possibility that SMR probes not only the spin part but rather the total the angular momentum this would not influence the position of the sharp resistance transition and the hysteresis. Therefore, whatever component contributes to the signal, the hysteresis width would be the same for both the spin and the orbital component.

Comparison of the model to the field sweep along $[110]$ presented in Fig.~\ref{fig:spinflopsminus} reveals good qualitative agreement. The modeled SMR marked as a red line exhibits the similar behavior to the measurements. The model (red solid line in Fig.~\ref{fig:spinflopsminus}) also reproduces the monotonic increase of the SMR that can be due to spin canting in high fields. It should be noted that the quantitative difference between the model and data can be explained by the fact that we deal with a domain structure and only part of the domains are initially oriented parallel to the field. Therefore, the magnitudes of the abrupt resistance change at $7\,\text{T}$ and the progressive resistance increase for $B>7\,\text{T}$ cannot be easily captured by the simple macrospin model.

\begin{figure}
\includegraphics[width=1\linewidth]{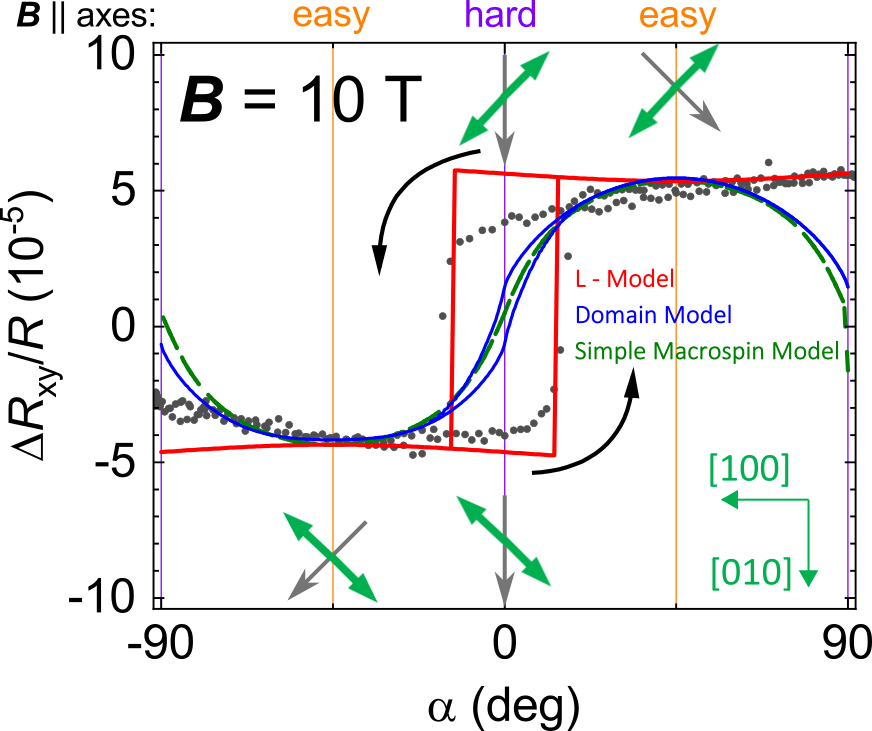}
\caption{\label{fig:10T} The angular dependence of magnetoresistance for the CoO structure collected at $B=10\,\text{T}$ is depicted by the black dots. It is expressed as relative changes of the resistance as a function of an angle $\alpha$ that $\bf{B}$ forms with the [010] crystalline direction. Green (gray) arrows represent the spin (magnetic field) orientation with respect to crystalline axes (coordinates in bottom right corner). Black arrows depict the angle sweeping direction. The green dashed line displays modeled behavior of the ADMR with $B_\text{sf}=7\,\text{T}$. Introduction of the new energy term into the model with $\xi=-4.02$ can reproduce the hysteresis as illustrated by the red continuous curve. The blue line represents the domain model.}
\end{figure}

\begin{figure}
\includegraphics[width=1\linewidth]{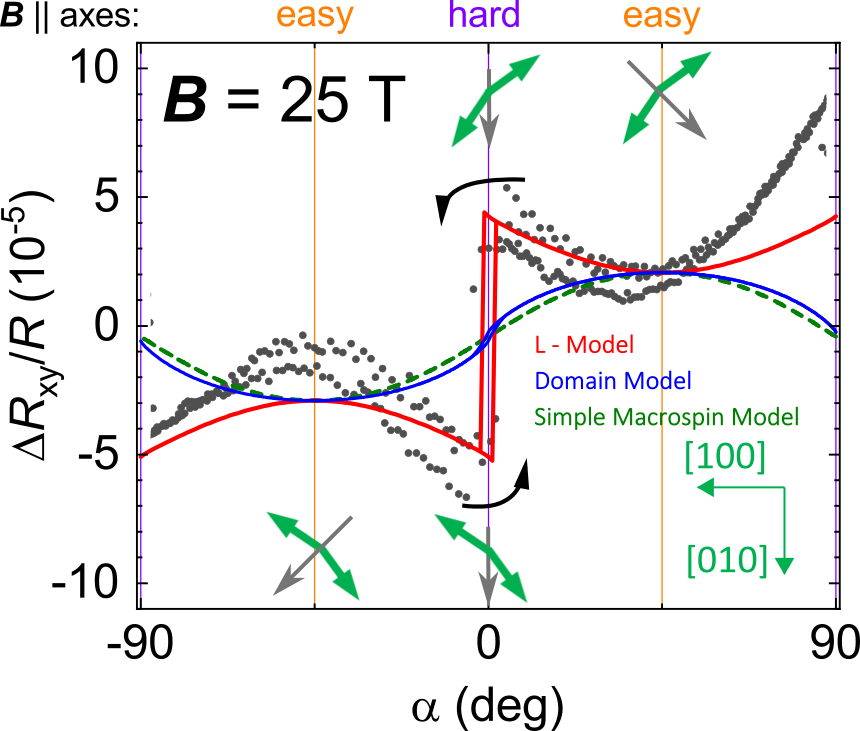}
\caption{\label{fig:25T} The angular dependence of the magnetoresistance for $B=25\,\text{T}$. Experimental results (black points) are compared to the macrospin models with $\xi=0$ (green dashed line) and $\xi=-4.02$ (red continuous curve) and the domain model (blue line). The corresponding spins (magnetic field) orientation with respect to the crystalline axes with the coordinates in bottom right corner are symbolized by the thick green (gray) arrows.}
\end{figure}

Despite a good correlation between the model and the experimental data for the field along an easy direction, we find that the model cannot describe the measurements for the magnetic field along a hard axis \cite{supplement}. To clarify it, we perform an angular dependent magnetoresistance (ADMR) measurements at different magnetic fields and show the data for $B = 10\,\text{T}$ and $B = 25\,\text{T}$ in Fig.~\ref{fig:10T} and in Fig.~\ref{fig:25T}, respectively. These angle-dependent measurements determine how $\bf{n}$ evolves upon gradually changing the magnetic field direction from an easy to a hard axis. In ADMR, the magnetic field vector has a fixed magnitude but it rotates with respect to the crystalline axes. The resistance is recorded as a function of an angle $\alpha$ between the magnetic field and [010], which is a hard axis. Above the $B_{\text{sf}}=7\,\text{T}$, $\bf{n}$ is expected to follow the direction perpendicular to $\bf{B}$, which is high enough to overcome the in-plane anisotropy \cite{Geprags2020JAP}. This should be reflected in the SMR signal that is shown as the green dashed lines in Fig.~\ref{fig:10T} and in Fig.~\ref{fig:25T}, which is the result of the macrospin model calculation with experimentally determined $B_\text{sf}$.

\begin{figure}
\includegraphics[width=1\linewidth]{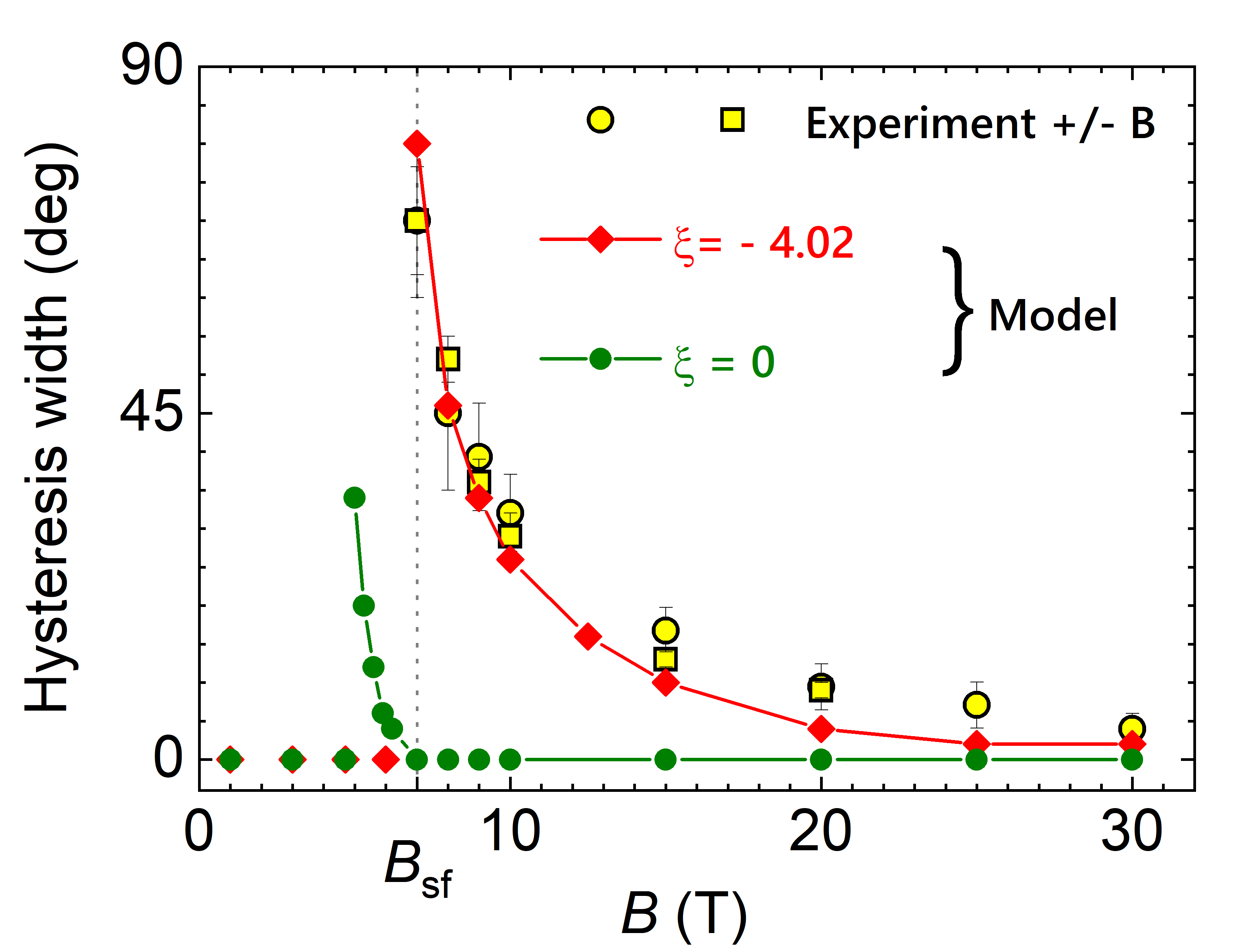}
\caption{\label{fig:hysts}Summary of the hysteresis width observed in ADMR as a function of the magnetic field (yellow squares and circles). The macrospin model with $\xi=0$ (green circles) predicts the occurrence of a hysteresis only slightly below $B_\text{sf}$, whereas setting $\xi=-4.02$ (red diamonds) yields good agreement with the measurements. Lines are guides for the eye.}
\end{figure}

Remarkably, the experiments show a completely different SMR dependence. When the magnetic field changes the orientation from $\alpha = 90^\circ$ to $\alpha = - 90^\circ$ and back, the electrical signal follows the same trend except for the region close to $\alpha=0^\circ$. In the vicinity of $\alpha=0^\circ$, the point where the magnetic field points along the hard axis, the SMR shows a hysteresis loop (Fig.~\ref{fig:10T} and Fig.~\ref{fig:25T}). At higher magnetic fields, the hysteresis becomes narrower. The hysteresis can be characterized by its width and plotted as a function of $\bf{B}$ (Fig.~\ref{fig:hysts}). Then, it can be clearly seen that hysteresis loops appear in the whole range of tested magnetic fields between $B_\text{sf}$ up to $30\,\text{T}$. The hysteresis loops are absent above $T_\text{N}$ and thus they can be attributed to the antiferromagnetic behavior of CoO. Below $B_\text{sf}$, no abrupt resistance changes are detectable and only slight residual hysteretic behavior around $\alpha=0$ can be observed that can be due to possible domain structure in the system \cite{supplement}. The ADMR measurements reveal that for $\mathbf{B} \parallel [010]$ and $B > B_\text{sf}$ different states of the AF are possible depending on the history and the system behaves as if it is always below the spin-flop field. The experimental results can be understood as the N\'eel vector lagging behind the magnetic field due to the strong in-plane anisotropy. However, such effect is expected only below $B_\text{sf}$, but should be absent above the spin-flop field in our current understanding of the interplay between Zeeman energy, antiferromagnetic exchange, and anisotropy energy, where $\mathbf{n}$ should follow the direction perpendicular to the magnetic field \cite{Geprags2020JAP}.

First, we check whether the existence of a domain structure can explain the experimental observations. We consider a set of N\'eel vectors that experience different anisotropy fields which reflects possible inhomogeneity of the layer that can result in antiferromagnetic domains. We simulate an ADMR measurement for each case separately using equation~(\ref{eqn:1}) and average obtained results assuming a Gaussian distribution of the spin-flop fields centered around $7\,\text{T}$ \cite{supplement}. We refer to this as a domain model \cite{supplement}. As can be seen in the blue line in Fig.~\ref{fig:10T} such a procedure can indeed result in appearance of a small hysteresis in ADMR above the spin-flop field. However, it completely fails to reproduce the abrupt transitions visible in strong magnetic fields (Fig.~\ref{fig:25T}) and is therefore not appropriate to explain this observation. 

Therefore, we consider another approach to interpret the experimental data. We assume that the magnetic field modifies the magnetic anisotropy of the antiferromagnet. We describe this contribution by introducing an additional, phenomenological energy term, relevant above $B_\text{sf}$:
\begin{equation}\label{eq-ksi}
E_a=\frac{\xi}{B_\text{sf}^2} B_xB_yn_xn_y 
\end{equation}
into the equation~(\ref{eqn:1}) with a coefficient $\xi$. It appears only if the magnetic field has nonzero components for both easy directions. Thus, it follows the symmetry of the experimental observations and can describe the unexpected hysteretic behavior of the AF when a strong magnetic field is parallel to a hard direction. For negative $\xi$, such contribution enhances the energy barrier between the states with different orientations of the Néel vector seen as an enhancement of the effective anisotropy field. We believe that this contribution originates from the field-induced tilting of the orbital momentum from the easy axis; and hence we call it L-model.
 
The existence of the hysteresis loop in SMR can be reproduced using the L-model with $\xi=-4.02$ as demonstrated by the solid red lines in Fig.~\ref{fig:10T} and Fig.~\ref{fig:25T}. Good agreement of the hysteresis width between the model and experiments is obtained for all magnetic fields tested experimentally (Fig.~\ref{fig:hysts}). Slight deviations from the model that can be seen in the experimental data in Fig.~\ref{fig:10T} as a rounded shape of the hysteresis loop and a difference in the curvature near $\alpha = \pm 45^{\circ}$ can be interpreted either as a signature of a domain structure and magnetic inhomogeneity of the system (these are also the features of the blue, domain model curve) or the contribution of the angular momentum into the SMR signal.  For the model with $\xi =0$, which corresponds to the original equation~(\ref{eqn:1}) and constant magnetic anisotropy, no hysteresis above the $B_\text{sf}$ is expected as indicated by green dashed lines in Fig.~\ref{fig:10T} and~\ref{fig:25T} as well as green circles in Fig.~\ref{fig:hysts}.

A likely explanation for the observed hysteretic behavior and the additional energy term  $E_a$ (equation~\ref{eq-ksi}) is the unquenched orbital momentum that is known to be present in CoO and has a strong contribution to the magnetocrystalline anisotropy in this material \cite{Kanamori1957I, Kanamori1957II}. It has also been observed to manifest itself in the magnon spectrum \cite{Satoh2017}. Furthermore, the orbital part of the angular momentum $L$ has already been shown to determine the angular magnetoresistance of CeSb, which is a 4f monopnictide with unquenched $L$ \cite{Xu2019}. In general, it is observed that the orbital part can induce a wide variety of transport phenomena \cite{Go2021}, which belong to the most recent discipline of orbitronics.

In our study, we restrict our discussion to the simplest case to qualitatively justify the orbital momentum dependence on the magnetic field and its influence on the anisotropy. First, we note that spin--orbit coupling induces an antiparallel alignment of the effective $L=1$  orbital angular momenta $\mathbf{L}_1$ and $\mathbf{L}_2$  to the corresponding magnetic spins for each sublattices \cite{Satoh2017}. The external magnetic field $B\le B_\text{sf}$ aligned parallel to the magnetic spins does not change this configuration; the orbital angular momenta are aligned antiparallel to spins with $L_{xj}=\pm 1$ (where the quantization axis $x$ is parallel to the easy axis). However, the magnetic field, oriented generically in the $xy$- plane, induces mixing of the states with the projections $L_{jx}=\pm1$, and  $0$. This mixing, in turn, results in a nonzero expectation values of  quadrupolar variables, $\langle \hat{L}_{jx}\hat{L}_{jy}\rangle$ ($j=1,2$) and additional spin anisotropy $K^\mathrm{ha}_\mathrm{an}A_{xy}(\mathbf{B})n_xn_y$ with $A_{xy}(\mathbf{B})=\sum_{j=1,2}\langle \hat {L}_{jx} \hat {L}_{jy}\rangle$.

To calculate the quantum states of the orbital momenta we introduce the hamiltonian \cite{Satoh2017}:
\begin{equation}\label{eq-hamiltonian-reduced}
\hat{\mathcal{H}}=\sum_{j=1,2}\left[-K^L\left(\hat{\mathbf{L}}_j\mathbf{e}_j\right)^2+\lambda\mathbf{S}_j\hat{\mathbf{L}}_j+2\mu_B\mathbf{B}\hat{\mathbf{L}}_j\right],
\end{equation}
where $K^L$ is anisotropy of the angular momentum, the constant $\lambda$ parametrizes spin-orbit coupling, $\mu_B$ is the Bohr magneton. We assume that the orientation of the N\'eel vectors is fixed along $x$ and  the local quantization axes $\mathbf{e}_j$ are parallel/antiparallel to the N\'eel vector ($\mathbf{e}_1\uparrow\downarrow\mathbf{e}_2\uparrow\uparrow\hat{x}$).

\begin{figure}
\includegraphics[width=1\linewidth]{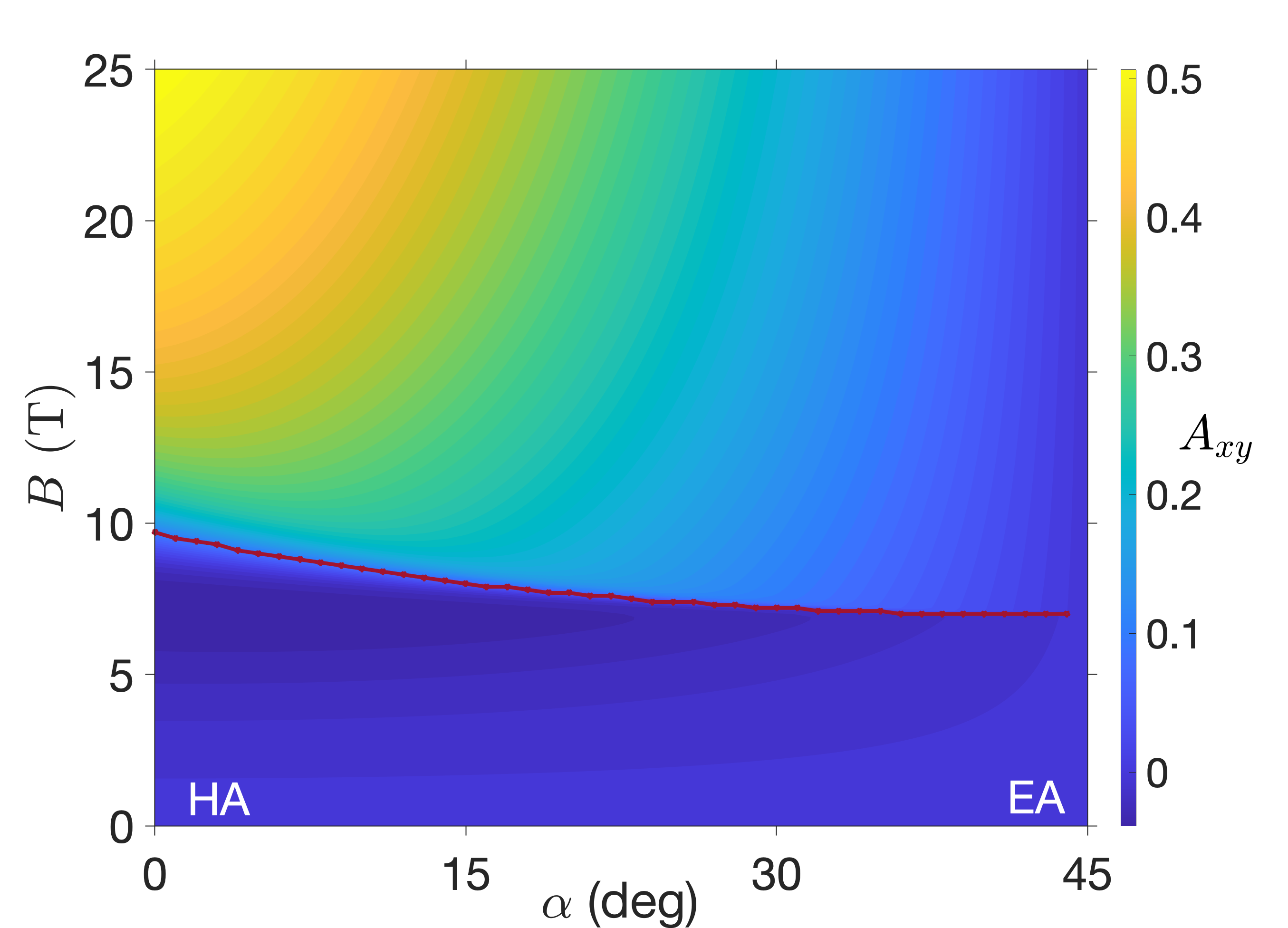}
\caption{\label{fig:L}The value of the field-induced anisotropy $A_{xy}(\mathbf{B})\equiv\sum_{j=1,2}\langle \hat{L}_{jx}\hat{L}_{jy}\rangle$ (color code, dimensionless) as a function of the orientation and the magnitude of the magnetic field. The star line shows the value of the threshold field $B_\mathrm{th}(\alpha)$ below which the field-induced anisotropy is negligible. The HA and EA denote a hard and an easy axis, respectively.}
\end{figure}
Figure~\ref{fig:L} shows the field dependence of $A_{xy}(\mathbf{B})$  (see supplementary material for more details). Below the critical field (red line), $A_{xy}(\mathbf{B})$ is close to zero, as the  quadrupoles $\langle \hat{L}_{1x}\hat{L}_{1y}\rangle$ and $\langle \hat{L}_{2x}\hat{L}_{2y}\rangle$ have opposite signs and almost compensate each other. However, close to the critical field value, the magnetic field is large enough to align both orbital momenta parallel to the magnetic field, and both quadrupoles have the same sign. This induces a rapid increase of $A_{xy}(\mathbf{B})$ above the critical field with the maximum anisotropy value achieved for the field parallel to the hard axis. There is a nonlinear dependence of $A_{xy}(\mathbf{B})$  on the magnetic field \cite{supplement} above the critical field, which we substitute with the $B^2$ relation for simplicity in our phenomenological energy term (equation~\ref{eq-ksi}). The value of the critical field scales with spin--orbit coupling $\lambda$ and corresponds to the spin-flop field $B_\mathrm{sf}$ if $\mathbf{B}$ is parallel to an easy axis. Field-induced rotation of orbital states can induce additional strain $u_{xy}\propto A_{xy}$ that through a magnetoelastic mechanism contributes into effective magnetic anisotropy.  We conjecture that a direct experimental confirmation of such behavior of the angular momentum could be performed, for example, by studying the magnetic field induced frequency shift of the magnon modes with optical excitation. We also notice a similarity of the considered case to the Pashen--Back effect, in which the magnetic field is much stronger than the spin-orbit coupling of a system \cite{LL}.

To summarize, we report the observation of the hysteresis loops in the angular dependence of magnetoresistance of a thin film of CoO with adjacent Pt layer. The hysteretic behavior is unexpectedly present above the spin-flop field and persists up to the highest tested magnetic fields ($30\,\text{T}$). It can be interpreted by the dependence of the magnetic anisotropy on the external magnetic field. The unquenched orbital angular momentum is a likely reason for the observed effects. These findings highlight the role of the anisotropy variations induced by the magnetic field and the role of the unquenched orbital momentum in the physics of antiferromagnets and their potential applications. The anisotropy dependence on the magnetic field may modify the expected equilibrium spin state of antiferromagnets in electrical experiments that strongly rely on the generation of the effective magnetic field.

We would like to thank Rembert Duine, Reinoud Lavrijsen, Michał Baj and Tomasz Dietl for helpful discussions. We acknowledge Pim Lueb for VSM-SQUID measurements. We acknowledge support from HFML-RU, member of the European Magnetic Field Laboratory (EMFL). Sample fabrication was performed using NanoLabNL facilities. The research was funded by the Dutch Research Council (NWO) under Grant 680-91-113. O.G. acknowledges support from the Alexander von Humboldt Foundation, the ERC Synergy Grant SC2 (No. 610115), and the Deutsche Forschungsgemeinschaft (DFG, German Research Foundation) - TRR 173 – 268565370 (project B12). This research was funded in part by National Science Centre, Poland 2021/40/C/ST3/00168. For the purpose of Open Access, the author has applied a CC-BY public copyright licence to any Author Accepted Manuscript (AAM) version arising from this submission.

\nocite{*}
\bibliography{GrzybowskiHystereticv8}

\end{document}